\documentclass[pra,twocolumn,aps]{revtex4}
\usepackage{amsmath}

\input{tcilatex}
\usepackage{graphicx}

\begin{document}

\title{A renormalized Hamiltonian approach to a resonant valence bond
wavefunction$^{1}$}
\author{F C Zhang, C Gros, T M Rice and H Shiba$^{2}$}
\address{Theoretische Physik, ETH-H$\overset{..}{o}$nggerberg, CH 8093
Zurich, Switzerland}

\begin{abstract}
The effective Hamiltonian of strongly correlated electrons on a square
lattice is replaced by a renormalised Hamiltonian and the factors that
renormalize the kinetic energy of holes and the Heisenberg spin-spin
coupling are calculated using a Gutzwiller approximation scheme. The
accuracy of this renormalization procedure is tested numerically and found
to be qualitatively excellent. Within the scheme a resonant valence bond
(RVB) wavefunction is found at half-filling to be lower in energy than the
antiferromagnetic state. If the wavefunction is expressed in fermion
operators, local SU(2) and U(l) invariance leads to a redundancy in the
representation. The introduction of holes removes these local invariances
and we find that a d-wave RVB state is lowest in energy. This state has a
superconducting order parameter whose amplitude is linear in the density of
holes.
\end{abstract}

\maketitle

\footnotetext[1]{%
Published in Supercond. Sci. Technol. 1 (1988) 36-46}

\footnotetext[2]{%
Permanent Address: Institute for Solid State Physics, \newline
University of Tokyo,Roppongi, Tokyo 106, Japan}

\section{Introduction}

Very soon after the discovery of high-$T_{c}$, superconductivity Anderson %
\cite{Anderson1} proposed that it was caused by a cooperative condensation
of carriers moving in a resonant valence bond (RVB) state of spins. Since
then this proposal has been studied extensively and the best account is in
Anderson's recent lecture notes \cite{Andseson 2}. Many other proposals have
been made (for a Review see \cite{Rice1}), and there have been questions
raised about the use of the simplified effective Hamiltonian derived from a
single-band Hubbard model in the atomic limit that forms the starting point
of Anderson's treatment. We shall not go into these questions here but just
point out that two of us (Zhang and Rice \cite{Zhang FC1}) recently gave an
explicit demonstration that a two-band model describing hybridised copper 3d
and oxygen 2p states can also be reduced to the same effective Hamiltonian
in an appropriate limit.

The effective Hamiltonian contains the strict local constraint which forbids
double occupancy of any site. This constraint is very difficult to handle
analytically. One of the most physically transparent methods to treat this
type of problem analytically has been the Gutzwiller approximation which
introduces a renormalization of the quantum mechanical expectation values by
a classical weighting factor \cite{Gutzwiller}. Such renormalisation can
then be incorporated into a Hamiltonian which may be treated by conventional
methods. This approach was used for the heavy fermion problem by Rice and
Ueda \cite{Rice2} and was shown to be equivalent to an optimal slave-boson
formulation by Kotliar and Ruckenstein \cite{Kotliar1}. In this paper we
will consider this renormilisation Hamiltonian method for the effective
Hamiltonian. Unlike the renormalised Anderson Hamiltonian studied by Rice
and Ueda \cite{Rice2}, in the present case the renormalised Hamiltonian
cannot be simply diagonalised and we must resort to a further mean-field
approximation. Mean-field approaches have been considered by many authors %
\cite{Baskaran1},\cite{Ruckenstein},\cite{Kotliar2}. Here we choose to
formulate the problem in terms of a variational wave-function. This has
several advantages. First it shows us that a consistent mean-field theory
must be formulated in terms of two expectation values i.e. one must include
particle-hole amplitudes of the form $\left\langle c^{+}c\right\rangle $ in
addition to particle-particle amplitudes of the form $\left\langle
c^{+}c^{+}\right\rangle $. This point has been recently realised by others
as well \cite{Suzumura}$_{+}^{+}\footnotetext[3]{$_{+}^{+}$In [8], Baskaran
and co-workers considered the term $\left\langle c^{+}c\right\rangle $as
well. However they set it equal to zero in their calculations.}$. The
coupled equations to minimise the energy have a wide class of degenerate
solutions at half-filling. Secondly a wave-function formulation is suited to
examining the role of the redundancy in the fermion representation which is
not present in the spin representation. At half-filling, this redundancy
which has its origin in the reduction from 4 to 2 degrees of freedom per
site as one goes from fermion to spin representation, appears as a local
particle-hole ($SU(2)$) and gauge ($U(1)$) invariance. The large degeneracy
of the mean-field description arises from this redundancy and it can be
shown that it corresponds to the same state in the spin representation.
Further the appearance of coherence in the fermion representation is
illusory so that there can be no true phase coherence as stressed by
Baskaran and Anderson \cite{Baskaran2}. Thirdly, this formulation allows a
direct comparison with the variational Monte Carlo (VMC) results. This
allows us on the one hand to check the validity of the renormalised
mean-field theory and on the other hand it gives us more insight into the
numerical VMC results. Both qualitatively and even quantitatively good
agreement is found; for example both point to a d-wave paired state as the
most stable and a true superconductivity order parameter which vanishes at
half-filling and grows linearly in the deviation from half-filling. The
largest discrepancy occurs for the antiferromagnetic state which, within
this scheme is higher in energy than the RVB state, contrary to the VMC
results.

Our treatment is essentially limited to zero temperature and the extension
to finite temperature will be non-trivial. Some discussion of the problems
of calculating excitation energies is given. In particular there are two
energy scales of excitations given by the gauge coherence energy (determined
by the kinetic energy) and the magnetic coherence energy respectively.
Anderson\cite{Andseson 2} has emphasised this splitting of the charged
excitations (holons) and spin excitations (spinons).

\section{The model and the renormalised Hamlltonlan}

We study the Hubbard model on a square lattice. In the limit of large
on-site Coulomb repulsion $U$ and at one-half, or, slightly less, filling
the Hubbard Hamiltonian can be transformed to the form%
\begin{eqnarray}
H &=&H_{t}+H_{S}  \TCItag{1}  \label{Equation1} \\
H_{t} &=&-t\underset{\left\langle i,j\right\rangle ,\sigma }{\sum }%
c_{i,\sigma }^{+}c_{j,\sigma }+HC  \notag \\
H_{S} &=&J\underset{\left\langle i,j\right\rangle }{\sum }S_{i}\cdot S_{j} 
\notag
\end{eqnarray}%
with the local constraint the number of electrons on any site $\leq $1. This
transformation has a long history, and has been used by \cite{Kohn}, amongst
others. In (\ref{Equation1}) $H_{t}$, and $H_{s}$, are the kinetic and
magnetic energies respectively and $\left\langle i,j\right\rangle $
represent the nearest-neighbour pairs. $S_{i}$ are the spin $=\frac{1}{2}$
operators and $J=\frac{4t^{2}}{U}$. We neglect terms which are higher order
in the small parameters $t/U$ and the hole concentration $\delta $ ($=1-n$
where $n$ is the electron concentration).

Since the high-$T_{c}$, superconducting materials show strong
antiferromagnetic (AF) spin correlations \cite{Vaknin} it is believed that
this model contains the essential physics for the high-$T_{c}$,
superconductivity \cite{Anderson1},\cite{Zhang FC1}.

To study the ground state and the excited states of (\ref{Equation1}), we
use a projected BCS trial wave-function as suggested by Anderson \cite%
{Anderson1} for a RVB state:%
\begin{equation}
\left| \varphi \right\rangle =P_{d}\left| \varphi _{0}\right\rangle  \tag{2}
\label{Equaiton2}
\end{equation}%
\begin{equation}
\left| \varphi _{0}\right\rangle =\underset{k}{\prod }(u_{k}+v_{k}c_{k%
\uparrow }^{+}c_{-k\downarrow }^{+})\left| 0\right\rangle  \tag{3}
\label{Equaiton3}
\end{equation}%
where the Gutzwiller projection operator $P_{d}=\underset{i}{\prod }%
(1-n_{i\uparrow }n_{i\downarrow })$ and $\left| 0\right\rangle $ is the
vacuum state. $u_{k}$ and $v_{k}$ are the variational parameters satisfying
the normalization condition for $\left| \varphi _{0}\right\rangle $: $\left|
u_{k}\right| ^{2}+\left| v_{k}\right| ^{2}=1$.

Some special forms of (\ref{Equaiton2}) have recently been studied
numerically. Using the VMC technique, which treats the projection operator
exactly \cite{Horsch}, \cite{Gros1}, \cite{Gros2}, \cite{Yokoyama1}, \cite%
{Yokoyama2}, the energies of these states have been numerically calculated.
It is found that in the square lattice, the projected Fermi liquid state
(i.e. the state with $u_{k}v_{k}^{\ast }=0$) is unstable against d-wave
pairing \cite{Gros1}. At half-filling the energy of the d-wave state is
found \cite{Gros2} to be very close to the ground-state energy extrapolated
from the exact small system calculations \cite{Oitmaa}. In contrast to the
extrapolated exact small system calculations, the d-wave trial wave-function
has no long range antiferromagnetic order \cite{Gros2} and may therefore be
viewed as an example of a quantum spin liquid. A VMC study of
superconductivity has been made independently by Yokoyama and Shiba \cite%
{Yokoyama2}. They also concluded a possibility of a d-wave superconductivity.

The projected BCS wave-function is a natural generalization of the usual BCS
state to strongly correlated systems. The projection operator, however,
makes difficulties for an analytic approach. In this paper we shall use a
renormalised Hamiltonian approach to treat the projection operator, and
systematically investigate the state (\ref{Equaiton2}), carrying out
explicitly the variational procedure. In this approach following Gutzwiller %
\cite{Gutzwiller} the effect of the projection operator on the doubly
occupied sites is taken into account by a classical statistical weighting
factor which multiplies the quantum coherent result calculated with $\left|
\varphi _{0}\right\rangle $. A clear description of the method has been
given by Vollhardt \cite{Volihardt}. The hopping energy and the spin-spin
correlation of the nearest neighbour sites in the state $\left| \varphi
\right\rangle $ are related to those in the state $\left| \varphi
_{0}\right\rangle $ by%
\begin{eqnarray*}
\left\langle c_{i\sigma }^{+}c_{j\sigma }\right\rangle &=&g_{t}\left\langle
c_{i\sigma }^{+}c_{j\sigma }\right\rangle _{0} \\
\left\langle S_{i}\cdot S_{j}\right\rangle &=&g_{S}\left\langle S_{i}\cdot
S_{j}\right\rangle _{0}
\end{eqnarray*}%
where $\left\langle A\right\rangle _{0}$ , and $\left\langle A\right\rangle $
are the expectation values in the states $\left| \varphi _{0}\right\rangle $%
, and $\left| \varphi \right\rangle $ respectively. The renormalization
factors $g_{t}$ and $g_{S}$ are determined by the ratios of the
probabilities of the corresponding physical processes in the states $\left|
\varphi \right\rangle $ and $\left| \varphi _{0}\right\rangle $. In figure 1
we illustrate the possible hopping processes in these two states. The
probability of such a process in the state $\left| \varphi \right\rangle $ is%
\begin{equation*}
\lbrack n_{j\uparrow }(1-n_{i})n_{i\uparrow }(1-n_{j})]^{1/2}
\end{equation*}%
while that in the state $\left| \varphi _{0}\right\rangle $ is%
\begin{equation*}
\lbrack n_{j\uparrow }(1-n_{i\uparrow })n_{i\uparrow }(1-n_{j\uparrow
})]^{1/2}
\end{equation*}%
$n_{i\sigma }$ are the average electron occupation numbers, ($n_{i}=\underset%
{\sigma }{\sum }n_{i\sigma }$) which are the same in the states $\left|
\varphi \right\rangle $ and $\left| \varphi _{0}\right\rangle $, because of
the spin symmetry of the wave-functions. This leads to the result \cite%
{Volihardt} 
\begin{equation}
g_{t}=2\delta /(1+\delta )  \tag{4a}  \label{Equation4a}
\end{equation}


\begin{figure}
\includegraphics[width=3.25in]{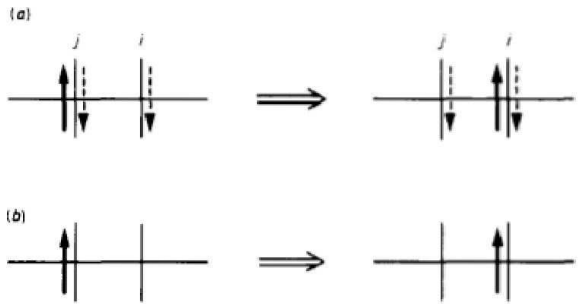}
\caption{\textit{The possible hopping
processes (a) in the non-projected pairing state (\ref{Equaiton3}) and (b)
in the projected BCS state (\ref{Equaiton2}). The spins with broken arrows
are optional in the (a) configurations.} }
\end{figure}

\FRAME{ftbpFU}{3.186in}{1.0767in}{0pt}{\Qcb{\textit{The spin exchange
process in the states (\ref{Equaiton2}) and (\ref{Equaiton3})}}}{\Qlb{Fig2}}{%
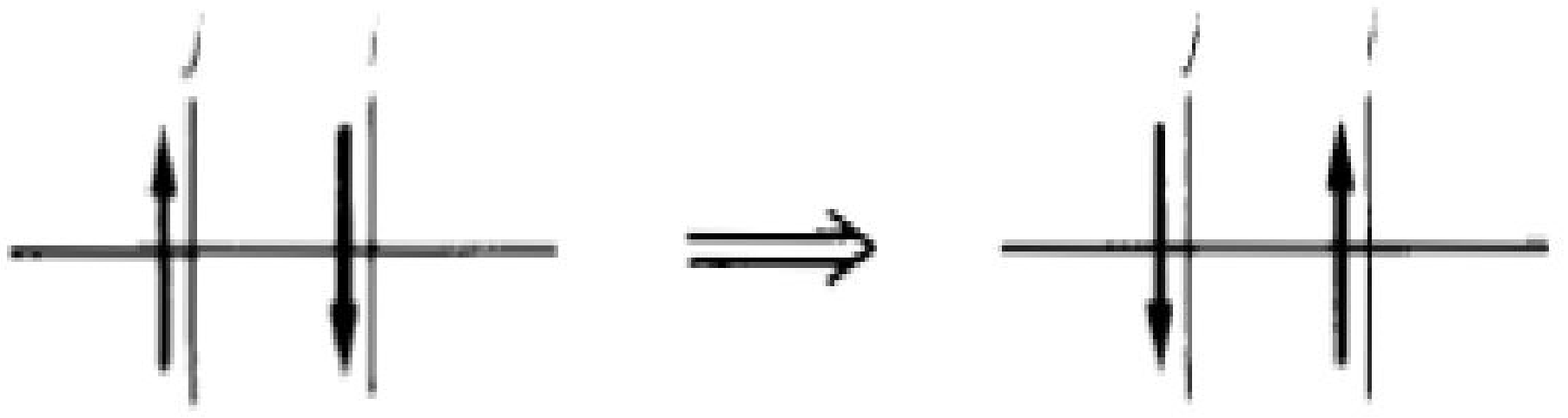}{\raisebox{-1.0767in}{\includegraphics[width=3.0in]{figure2.eps}}} \ \ \ 

To determine $g_{S}$, we consider the spin exchange process shown in figure
2. The spin-spin interaction occurs only when both sites are singly
occupied. The probability for such a process in the state $\left| \varphi
\right\rangle $ is $(n_{j\uparrow }n_{i\downarrow }n_{j\downarrow
}n_{i\uparrow })^{1/2}$, while in the state $\left| \varphi
_{0}\right\rangle $ it is $[n_{j\uparrow }(1-n_{j\downarrow })n_{i\downarrow
}(1-n_{i\uparrow })n_{j\downarrow }(1-n_{j\uparrow })n_{i\uparrow
}(1-n_{i\downarrow })]^{1/2}$. The same result is obtained for the $z$
component interaction $S_{i}^{z}S_{j}^{z}$. Thus one finds$^{+}%
\footnotetext[4]{$^{+}$In a systematic series expansion on $\delta $ and $%
J/t $, the higher-order terms of $\delta $ in (\ref{Equation4a},\ref%
{Equation4b}) should be dropped away to be consistent with the effective
Hamiltonian (\ref{Equation1}), where the higher-order terms are not
included. This, however, does not change the qualitative physics discussed
in this paper.}$%
\begin{equation}
g_{s}=4/(1+\delta )^{2}  \tag{4b}  \label{Equation4b}
\end{equation}%
It is important to realise that the projection operator greatly enhances the
spin-spin correlations. To further illustrate this point, we list in table 1
all the possible two-site states together with their weights and the
contributions to the spin-spin correlation in the half-filled case.\FRAME{%
ftbpFU}{3.3736in}{2.8167in}{0pt}{\Qcb{\textit{This table illustrates the
enhancement of the spin-spin Correlation in the projected BSC state }$\left| 
\protect\varphi \right\rangle $\textit{, equation (\ref{Equaiton2}), over
that in the BCS state }$\left| \protect\varphi \right\rangle _{0}$\textit{\ (%
\ref{Equaiton3}), at half-filling. The weight of the configurations actually
contributing \ to }$\left\langle S_{i}\cdot S_{j}\right\rangle $\textit{\
increases by a factor of four due to the projection. The configurations at
each site are denoted by 0 (empty state), }$\uparrow \downarrow $\textit{\
(doubly occupied state), and }$\protect\sigma $\textit{\ (singly occupied
state with spin }$\protect\sigma $\textit{) }}}{\Qlb{Table1}}{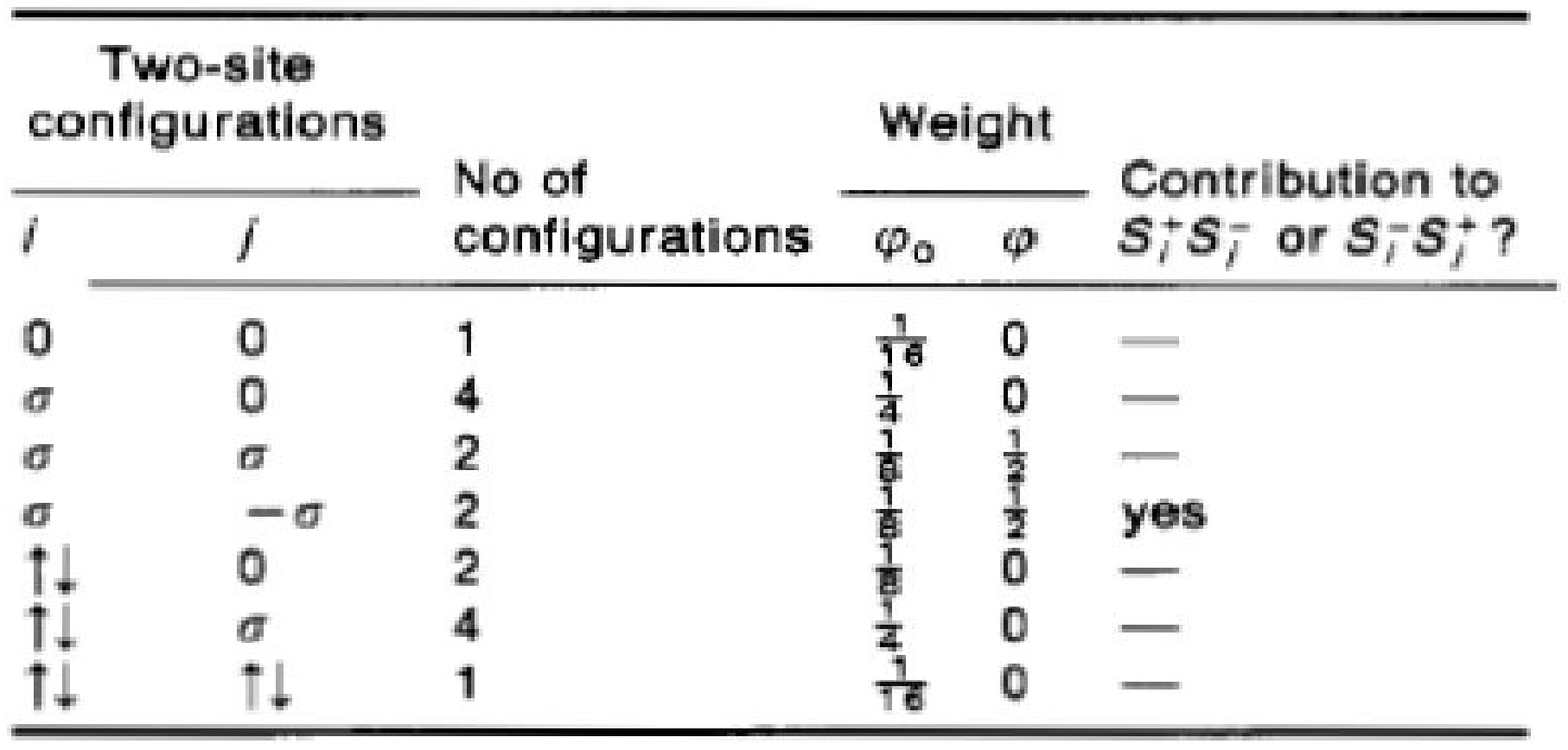}{%
\raisebox{-2.8167in}{\includegraphics[width=3.0in]{table1.eps}}}

Having determined the renormalisation factors, we can define a renormalised
Hamiltonian given by 
\begin{equation}
H^{^{\prime }}=g_{t}H_{t}+g_{S}H_{S}  \tag{5}  \label{Equation5}
\end{equation}%
The energy of the system in the state $\left| \varphi \right\rangle $ can be
evaluated as the expectation value of $H^{^{\prime }}$ in the state $\left|
\varphi _{0}\right\rangle $%
\begin{equation}
W=\left\langle H^{^{\prime }}\right\rangle _{0}  \tag{6}  \label{Equation6}
\end{equation}%
Equations (\ref{Equation4a},\ref{Equation4b})-(\ref{Equation6}) form the
basis of our renormalised Hamiltonian approach, which is analogous to the
approach used by Rice and Ueda \cite{Rice2} for the periodic Anderson model
with the difference that here we make a further mean-field approximation.
This is because in the context of the periodic Anderson Hamiltonian the most
important physical effect is the renormalisation of the f-level to the Fermi
surface and not the spin interaction, which would make an exact treatment of
the effective Hamiltonian impossible.

To justify this approach, we have carried out Monte Carlo calculations.
Figures 3-5 show the comparisons between the renormalised mean-field theory
and the essentially exact MC results for these wave-functions. The
quantitative agreement is within 5-15\%, while the qualitative agreement is
excellent for the wave-function (\ref{Equaiton2}).\FRAME{ftbpFU}{3.6504in}{%
3.3382in}{0pt}{\Qcb{\textit{A comparison between the renormalised mean-field
theory (RMF) (see \ref{Equation4a}, \ref{Equation4b}, \ref{Equation5}, \ref%
{Equation6}) and the Monte Carlo (MC) result for the kinetic energy }$%
\left\langle T\right\rangle $\textit{\ per hole in the projected BCS state (%
\ref{Equaiton2}). Both were calculated with a total number of sites }$%
N_{S}=82$\textit{\ and a number of holes }$N_{h}=8,16$\textit{. The
variational parameter }$\Delta $\textit{\ is related to the parameters of
the state (\ref{Equaiton2}) by }$\frac{v_{k}}{u_{k}}=\frac{\Delta _{k}}{%
\protect\varepsilon _{k}-\protect\mu _{0}+[\Delta _{k}^{2}+(\protect%
\varepsilon _{k}-\protect\mu _{0})^{2}]^{1/2}}$\textit{\ where }$\protect\mu %
_{0}$\textit{\ is a parameter, and }$\protect\varepsilon _{k}$\textit{\ is
given by (\ref{Equation8b}). In the d-wave pairing state, }$\Delta
_{k}=\Delta (\cos (k_{x})-\cos (k_{y}))$\textit{\ and in the s-wave state, }$%
\Delta _{k}=\Delta $\textit{. The full circles and squares are the MC
results for the s- and d-waves respectively. The dotted and broken curves
through the MC results are guides for the eyes. The second pair of dotted
and broken curves are the results from RMF for s- and d-waves respectively.}}%
}{\Qlb{Fig3}}{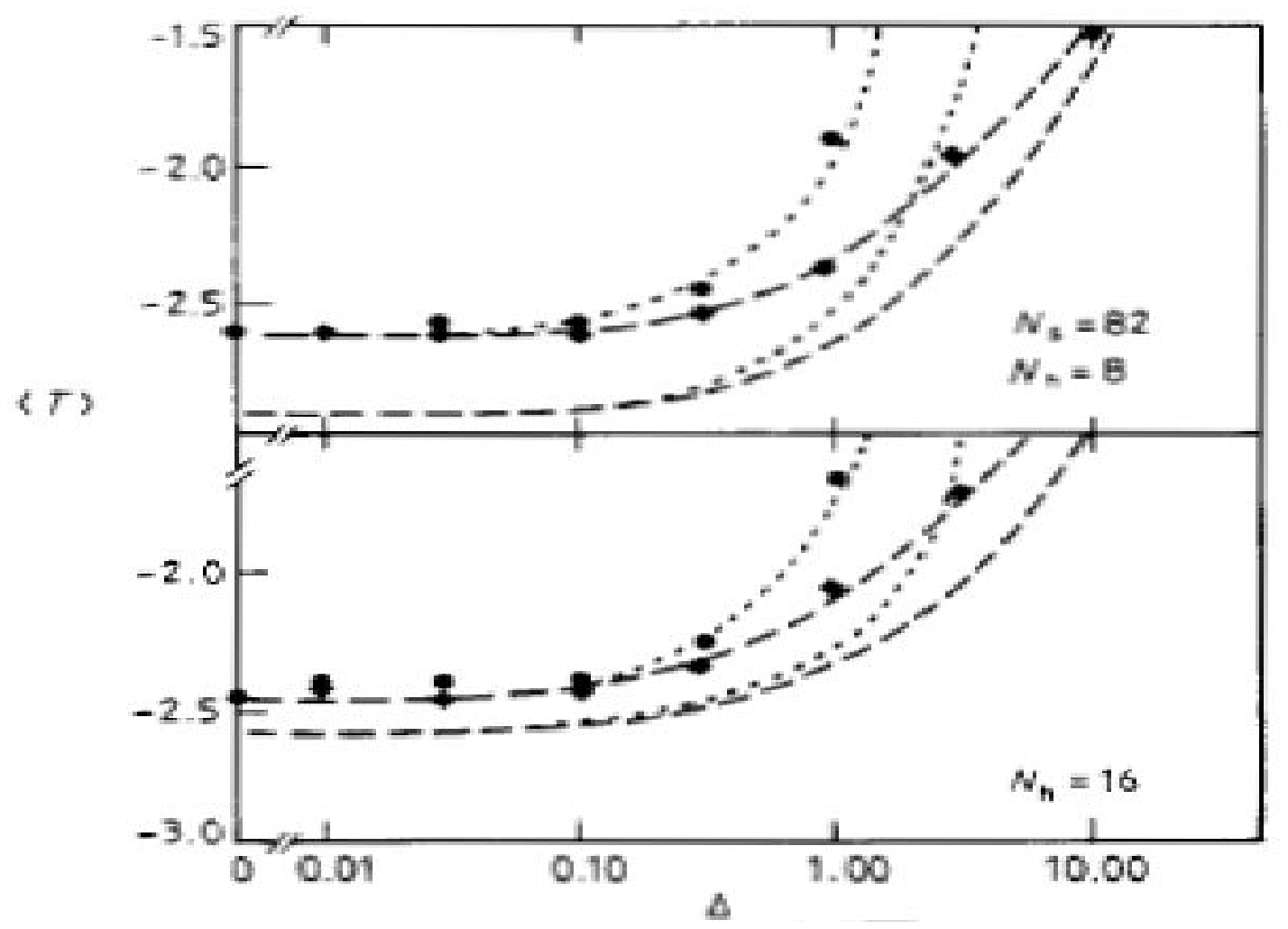}{\raisebox{-3.3382in}{\includegraphics[width=3.0in]{figure3.eps}}}\FRAME{ftbpFU}{3.3728in}{%
4.1321in}{0pt}{\Qcb{\textit{A comparison between the renormalised mean-field
theory (\ref{Equation4a}, \ref{Equation4b}, \ref{Equation5}, \ref{Equation6}%
) (RMF) and Monte Carlo (MC) results for the nearest neighbour spin-spin
correlation }$\left\langle S_{i}\cdot S_{j}\right\rangle $\textit{\ in the
projected BCS state (\ref{Equaiton2}). Both were calculated with a total
number of sites }$N_{S}=82$\textit{\ and a number of holes }$N_{h}=0,8,16$%
\textit{. The variational parameter }$\Delta $\textit{\ is related to the
parameters at the state (\ref{Equaiton2}) by }$\frac{v_{k}}{u_{k}}=\frac{%
\Delta _{k}}{\protect\varepsilon _{k}-\protect\mu _{0}+[\Delta _{k}^{2}+(%
\protect\varepsilon _{k}-\protect\mu _{0})^{2}]^{1/2}}$\textit{\ where }$%
\protect\mu _{0}$\textit{\ is a parameter and }$\protect\varepsilon _{k}$%
\textit{\ is given by (M). In the d-wave pairing state, }$\Delta _{k}=\Delta
(\cos (k_{x})-\cos (k_{y}))$\textit{\ and in the s-wave state }$\Delta
_{k}=\Delta $\textit{. The full circles and squares are the MC results for
the s- and d-waves respectively. The dotted and broken curves through them
are guides for the eyes. The second pair of dotted and broken curves are the
results from RMF for the s- and d-waves respectively.}}}{\Qlb{Fig4}}{%
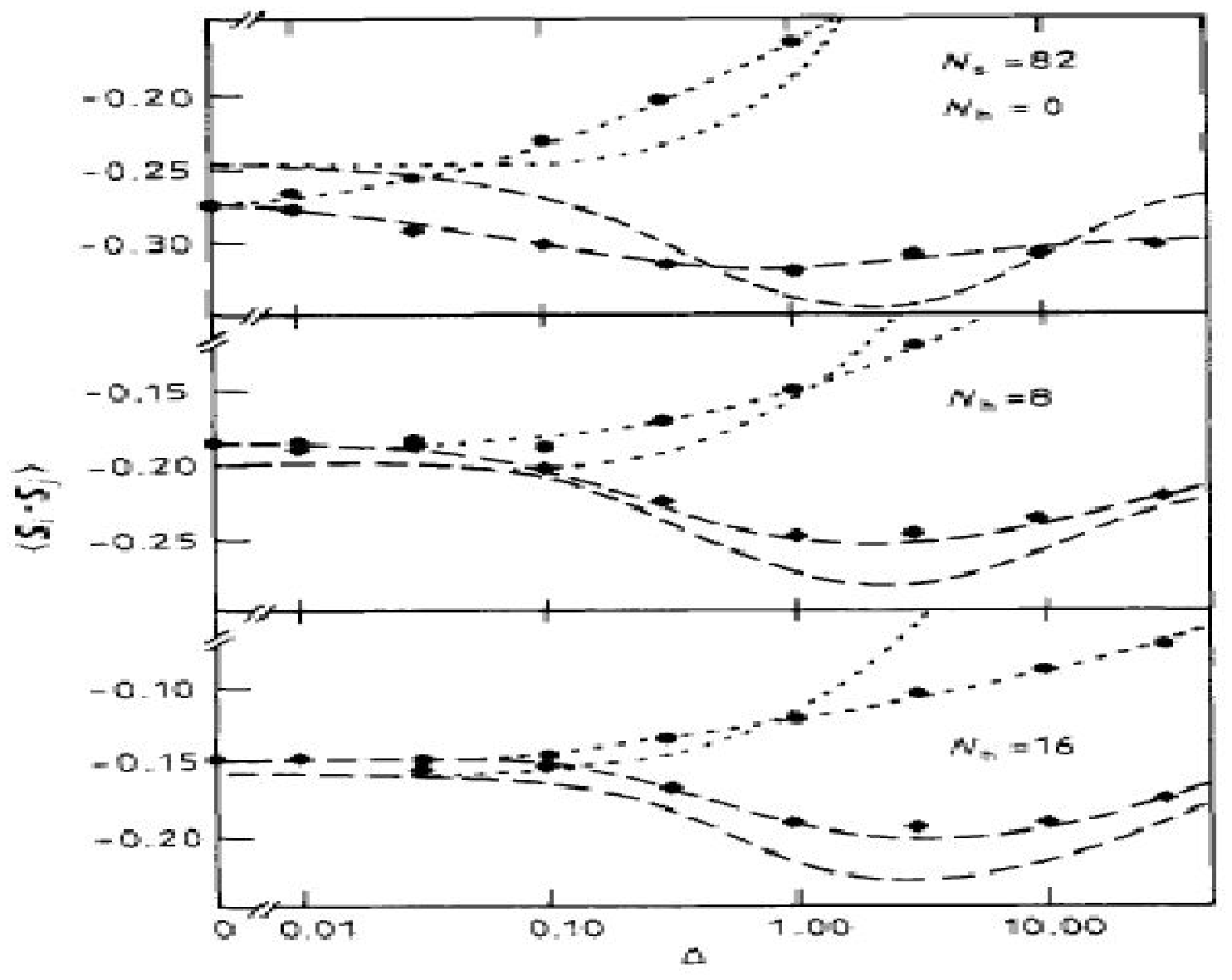}{\raisebox{-4.1321in}{\includegraphics[width=3.0in]{figure4.eps}}}\FRAME{ftbpFU}{3.122in}{2.469in%
}{0pt}{\Qcb{\textit{The nearest neighbour spin-spin correlation function }$%
\left\langle S_{i}\cdot S_{j}\right\rangle $\textit{\ as a function of
electron filling in the projected Fermi liquid state (}$u_{k}v_{k}=0$\textit{%
\ in (\ref{Equaiton2})). The renormalised mean-field theory (\ref{Equation4a}%
, \ref{Equation4b}, \ref{Equation5}, \ref{Equation6}) (RMF), broken curve,
agrees well with the Monte Carlo (MC), full circles, result in the entire
filling region. }}}{\Qlb{Fig5}}{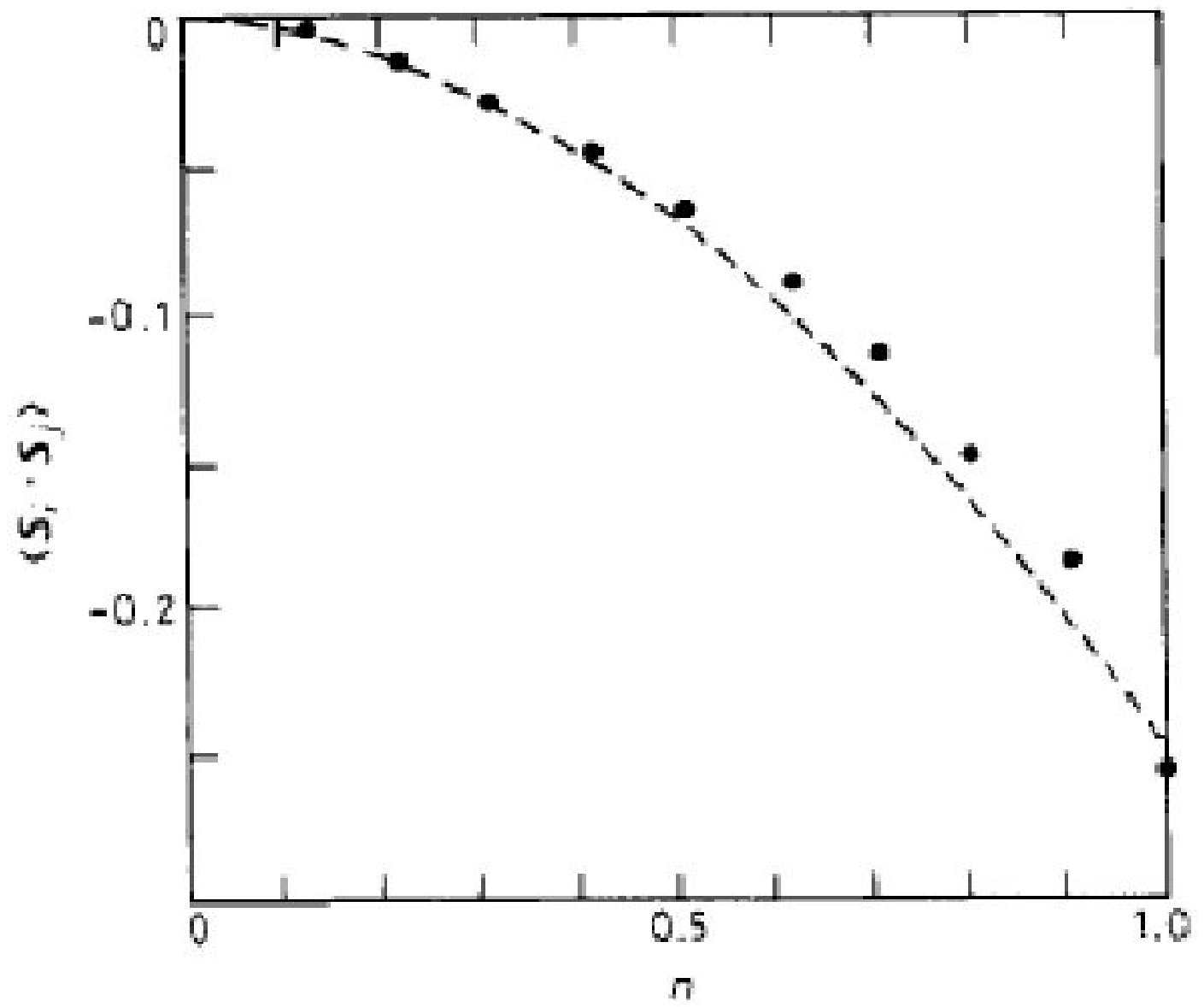}{\raisebox{-2.469in}{\includegraphics[width=3.0in]{figure5.eps}}}

After replacing the projection operator, the energy of the system can be
evaluated analytically. The variational task is to minimise $W$ in (\ref%
{Equation6}). This leads to coupled gap equations, which we will derive and
solve in the following sections.

\section{Gap Equation}

In this section we derive the gap equations for the projected BCS
wave-function within the renormalised Hamiltonian scheme described in
section2. We consider only the even-parity case, i.e., $u_{-k}v_{-k}^{\ast
}=u_{k}v_{k}^{\ast }$, and $\left| v_{-k}\right| ^{2}=\left| v_{k}\right|
^{2}$.

Evaluating (\ref{Equation6}), we obtain

\begin{eqnarray}
W &=&2g_{t}\underset{k}{\sum }\varepsilon _{k}\left| v_{k}\right| ^{2} 
\TCItag{7}  \label{Equation7} \\
&&+N_{S}^{-1}\sum_{k,k^{^{\prime }}}(V_{k-k^{^{\prime }}}(\left|
v_{k}\right| ^{2}\left| v_{k^{^{\prime }}}\right| ^{2}+u_{k}v_{k}^{\ast
}v_{k^{^{\prime }}}u_{k^{^{\prime }}}^{\ast })  \notag
\end{eqnarray}%
where $N_{S}$ is the total number of sites, and 
\begin{equation}
V_{k}=-\frac{3}{4}g_{S}J\gamma _{k}  \tag{8a}  \label{Equation8a}
\end{equation}%
\begin{equation}
\varepsilon _{k}=-t\gamma _{k}  \tag{8b}  \label{Equation8b}
\end{equation}%
\begin{equation}
\gamma _{k}=2(\cos k_{x}+\cos k_{y})  \tag{8c}  \label{Equation8c}
\end{equation}

Note that $\varepsilon _{k}$ and $V_{k}$ have the same functional form,
since $H_{S}$ is derived by kinetic exchange. The electron number operator $%
N=\underset{k,\sigma }{\sum }c_{k,\sigma }^{+}c_{k,\sigma }$ has expectation
value $\left\langle N\right\rangle =2\underset{k}{\sum }\left| v_{k}\right|
^{2}$.

Let $\mu $ be the chemical potential of the system, the quantity we want to
minimise is 
\begin{equation*}
\widetilde{W}=\left\langle H^{^{\prime }}-\mu N\right\rangle _{0}
\end{equation*}%
While minimising $\widetilde{W}$ with respect to $u_{k}$ and $v_{k}$ for
fixed $\mu $, one must remember that $H^{^{\prime }}$ is also a function of $%
\delta $ due to the renormalisation factors, and hence a function also of $%
v_{k}$. Carrying out this procedure, we find that 
\begin{eqnarray}
\left| u_{k}\right| ^{2} &=&\frac{1}{2}[1+\frac{\zeta _{k}}{E_{k}}] 
\TCItag{9a}  \label{Equaiton9a} \\
\left| v_{k}\right| ^{2} &=&\frac{1}{2}[1-\frac{\zeta _{k}}{E_{k}}] 
\TCItag{9b}  \label{Equaiton9b}
\end{eqnarray}%
and 
\begin{equation}
u_{k}v_{k}^{\ast }=\widetilde{\Delta }_{k}/2E_{k}  \tag{9c}
\label{Equation9c}
\end{equation}%
where%
\begin{equation}
E_{k}=(\zeta _{k}^{2}+\widetilde{\Delta }_{k}^{2})^{1/2}  \tag{9d}
\label{Equation9d}
\end{equation}

The parameters $\widetilde{\Delta }_{k}$and $\zeta _{k}$ are dimensionless,
and they are related to the particle-particle and particle-hole pairing
amplitudes respectively%
\begin{equation*}
\widetilde{\Delta }_{k}=\widetilde{\Delta }_{x}\cos (k_{x})+\widetilde{%
\Delta }_{y}\cos (k_{y})
\end{equation*}%
\begin{equation}
\zeta _{k}=\widetilde{\varepsilon _{k}}-\zeta _{x}\cos (k_{x})-\zeta
_{y}\cos (k_{y})  \tag{10}  \label{Equation10}
\end{equation}%
where 
\begin{equation*}
\widetilde{\varepsilon _{k}}=\frac{(g_{t}\varepsilon _{k}-\widetilde{\mu })}{%
\frac{3}{4}g_{S}J}
\end{equation*}%
and $\widetilde{\mu }$ is related to the chemical potential $\mu $ by 
\begin{equation}
\mu =\widetilde{\mu }-N_{S}^{-1}\left\langle \frac{\partial H^{^{\prime }}}{%
\partial \delta }\right\rangle _{0}  \tag{11}  \label{Equation11}
\end{equation}%
In (\ref{Equation10}) 
\begin{equation}
\widetilde{\Delta }_{\tau }=\left\langle c_{i\uparrow }^{+}c_{i+\tau
\downarrow }^{+}-c_{i\downarrow }^{+}c_{i+\tau \uparrow }^{+}\right\rangle
_{0}  \tag{12}  \label{Equation12}
\end{equation}%
\begin{equation}
\zeta _{\tau }=\sum_{\sigma }\left\langle c_{i\sigma }^{+}c_{i+\tau \sigma
}\right\rangle _{0}  \tag{13}  \label{Equation13}
\end{equation}%
with $\tau =x$ and $y$, $i+\tau $ denotes the NN of $i$ in the $\tau $
direction. Since we consider the even-parity case, $\zeta _{\tau }$ is real,
but $\widetilde{\Delta }_{\tau }$ can be complex. $\widetilde{\Delta }_{\tau
}$ and $\zeta _{\tau }$, satisfy the following coupled gap equations:%
\begin{equation*}
\widetilde{\Delta }_{k}=N_{S}^{-1}\sum_{k^{^{\prime }}}\gamma
_{k-k^{^{\prime }}}\frac{\widetilde{\Delta _{k^{^{\prime }}}}}{%
2E_{k^{^{\prime }}}}
\end{equation*}%
\begin{equation}
\zeta _{k}=\widetilde{\varepsilon _{k}}+N_{S}^{-1}\sum_{k^{^{\prime
}}}\gamma _{k-k^{^{\prime }}}\frac{\zeta _{k^{^{\prime }}}}{2E_{k^{^{\prime
}}}}  \tag{14}  \label{Equation14}
\end{equation}%
The first one is the same as the usual BCS gap equation. The second one
originates from the particle-hole correlation. From (\ref{Equation12}), it
is clear that $\widetilde{\Delta }_{k}$\ is related to the pairing in the
unprojected state $\left| \varphi _{0}\right\rangle $. It describes the
'smearing' of the pseudo-Fermi surface. However, $\widetilde{\Delta }_{k}$
is not the superconducting order parameter in the projected state $\left|
\varphi \right\rangle $ in our theory. $E_{k}$ turns out to be the
quasi-particle excitation energy (in units of $\frac{3}{4}g_{S}J$) in the
pairing state as we will show in section5.

The coupled gap equations (\ref{Equation14}) are the basic equations in our
approach. They can also be written in the $x$ and $y$ component form:%
\begin{equation*}
\widetilde{\Delta }_{\tau }=N_{S}^{-1}\underset{k}{\sum }\frac{\widetilde{%
\Delta _{k}}}{E_{k}}\cos k_{\tau }
\end{equation*}%
\begin{equation}
\zeta _{\tau }=-N_{S}^{-1}\underset{k}{\sum }\frac{\zeta _{k}}{E_{k}}\cos
k_{\tau }  \tag{15}  \label{Equation15}
\end{equation}%
The gap equations must be solved simultaneously with the hole concentration
equation, $\delta =N_{S}^{-1}\underset{k}{\sum }\frac{\zeta _{k}}{E_{k}}$.

Before we discuss the non-trivial solutions, we note that $\widetilde{\Delta
_{k}}=0$ is a trivial solution of the gap equations. This corresponds to the
projected Fermi-liquid state. In this case, $\zeta _{k}$ changes sign at the
Fermi surface. The parameters $\zeta _{x}=\zeta _{y}(=\zeta )$ are given by%
\begin{equation*}
\zeta =N_{S}^{-1}\underset{\zeta _{k}\leq 0}{\sum }\cos k_{x}+\cos k_{y}
\end{equation*}%
The volume of the Fermi sea is determined by the number of electrons. The
energy per site is 
\begin{equation*}
w=-4g_{t}t\zeta -\frac{3}{4}g_{S}J\zeta ^{2}
\end{equation*}%
In particular, $w=-48/\pi ^{4}J\simeq -0.49J$ in the half-filled case. It
will be shown in the next section that this trivial solution is unstable
against the pairing states with $\widetilde{\Delta _{k}}\neq 0$.

\section{Solutions to the gap equations---half-filled case}

At the half-filling,$\delta =0$, $\widetilde{\mu }=0$ and there is no
kinetic energy. We are interested in the possible lowest energy states of
the solution of (\ref{Equation14}). The total energy of the system has a
simple form in this case by use of (\ref{Equation7}) and (\ref{Equation14})%
\begin{equation*}
w=-\frac{3}{8}g_{S}J\underset{k}{\sum }E_{k}
\end{equation*}%
Therefore the lowest energy states correspond to the maximum value of $%
\underset{k}{\sum }E_{k}$. For this reason, we use an ansafz for $E_{k}$ to
examine the solutions of the gap equations$_{+}\footnotetext[5]{$_{+}$There
might be other solutions for the gap equations. We believe that theform of
the energy given by (\ref{Equation16}) givesthe lowest energy.}$%
\begin{equation}
E_{k}=C(\cos ^{2}(k_{x})+\cos ^{2}(k_{y}))^{1/2}  \tag{16}
\label{Equation16}
\end{equation}%
where $C$ is a parameter to be determined. Note that such a choice gives
only four point zeros for $E_{k}$. The gap equations then reduce to a single
equation, and we get 
\begin{equation*}
C=\frac{1}{2N_{S}}\underset{k}{\sum }(\cos ^{2}(k_{x})+\cos
^{2}(k_{y}))^{1/2}
\end{equation*}%
which has numerical value $C\approx 0.479$. The energy per site is%
\begin{equation*}
w=-\frac{3}{4}g_{S}JC^{2}=-0.688J
\end{equation*}%
This energy is much lower (about 20\%) than that of the projected
Fermi-liquid state found above. The parameter $\widetilde{\Delta }_{k}$
describes the pairing correlation in the renormalised Hamiltonian. Finite
values of $\widetilde{\Delta }_{k}$ indicate the binding of the electron
pairs in the pairing states (\ref{Equaiton3}).

We now determine the parameters $\widetilde{\Delta }_{\tau }$ and $\zeta
_{\tau }$, \ required for the choice of $E_{k}$ in (\ref{Equation16}). Using
(\ref{Equation9d}) and (\ref{Equation10}), we find that they should satisfy
the following simultaneous equations:%
\begin{equation*}
\zeta _{x}^{2}+\left| \widetilde{\Delta }_{x}\right| ^{2}=\zeta
_{y}^{2}+\left| \widetilde{\Delta }_{y}\right| ^{2}=C^{2}
\end{equation*}%
\begin{equation}
2\zeta _{x}\zeta _{y}+(\widetilde{\Delta }_{x}\widetilde{\Delta }_{y}^{\ast
}+HC)=0  \tag{17}  \label{Equation17}
\end{equation}

There is a wide class of parameters which satisfy the \ conditions (\ref%
{Equation17}). All the states in this class give the same expectation value
of the renormalised Hamiltonian $H^{^{\prime }}$ in (\ref{Equation5}).
Therefore at the half-filling, $H^{^{\prime }}$ has a large degeneracy of
ground states. For real $\widetilde{\Delta }_{x}$ and $\widetilde{\Delta }%
_{y}$, these states can be illustrated diagrammatically as shown in figure
6(a). A few examples of these states are

d-wave pairing :%
\begin{equation}
\widetilde{\Delta }_{x}=-\widetilde{\Delta }_{y}=\zeta _{x}=\zeta _{y}=C/%
\sqrt{2}  \tag{18a}  \label{Equaiton18a}
\end{equation}

d-wave density matrix:%
\begin{equation}
\widetilde{\Delta }_{x}=\widetilde{\Delta }_{y}=\zeta _{x}=-\zeta _{y}=C/%
\sqrt{2}  \tag{18b}  \label{Equaiton18b}
\end{equation}

chiral state:%
\begin{equation}
\widetilde{\Delta }_{x}=-i\widetilde{\Delta }_{y}=C,\text{ \ \ \ \ \ \ }%
\zeta _{x}=\zeta _{y}=0  \tag{18c}  \label{Equaiton18c}
\end{equation}

anisotropic state:%
\begin{equation}
\widetilde{\Delta }_{x}=\zeta _{y}=C,\text{ \ \ \ \ \ \ \ \ \ }\widetilde{%
\Delta }_{y}=\zeta _{x}=0  \tag{18d}  \label{Equaiton18d}
\end{equation}%
We remark that the d-wave density matrix state is different from the
extended s-wave state proposed by Baskaran, Zou and Anderson \cite{Baskaran1}%
. In their theory, the particle-hole amplitude $\left\langle
c^{+}c\right\rangle _{0}$ is not included, i.e. $\zeta _{x}=\zeta _{y}=0$.
Therefore their state has the same energy as the projected Fermi-liquid
state. The d-wave pairing state was studied numerically in \cite{Gros2} and %
\cite{Yokoyama1}, and the chiral state was discussed in \cite{Kotliar2}.
They belong to the solutions of the same gap equations in the present \
approach.\FRAME{ftbpFU}{3.3935in}{2.1205in}{0pt}{\Qcb{\textit{(a)
Diagrammatic illustration of the degenerate ground states for the
renormalised Hamiltonian (\ref{Equation5}) at half-filling. }$\protect\zeta =%
\protect\zeta _{x}\widehat{x}+\protect\zeta _{y}\widehat{y}$\textit{, }$%
\Delta =\widetilde{\Delta }_{x}\widehat{x}+\widetilde{\Delta }_{y}\widehat{y}
$\textit{\ with }$\protect\zeta _{\protect\tau }$\textit{, }$\widetilde{%
\Delta }_{\protect\tau }$\textit{\ given by (\ref{Equation12}) and (\ref%
{Equation13}). The full arrows represent the d-wave pairing state with }$%
\protect\zeta \perp \Delta $\textit{, and }$\left| \protect\zeta \right|
=\left| \Delta \right| $\textit{. All the states in (\ref{Equation17}) with
the real parameters }$\widetilde{\Delta }_{\protect\tau }$\textit{\ may be
obtained by rotating }$\protect\zeta $\textit{\ and }$\Delta $\textit{\
simultaneously by angles }$\protect\theta $\textit{, or reflecting the two
vectors about the }$\widehat{x}$\textit{\ axis. (b) An illustration of the \ 
}$SU(2)$\textit{\ gauge transformation. A state described by (}$\protect%
\zeta ^{^{\prime }}$\textit{,}$\Delta ^{^{\prime }}$\textit{) in (a) can be
obtained by a local }$SU(2)$\textit{\ from the state (}$\protect\zeta ,$%
\textit{\ }$\Delta $\textit{), under which }$c_{i,\protect\sigma }^{+}$%
\textit{\ at the four sites of the plaquette transform according to (\ref%
{Equation19}), with (}$\protect\alpha _{i}$\textit{,}$\protect\beta _{i}$%
\textit{) as denoted. The minus sign in front of the parentheses corresponds
to the states in (a) after a reflection about the }$\widehat{x}$\textit{\
axis. The transformation operator }$c_{i,\protect\sigma }$\textit{\ at other
lattice sites is determined by a translation.}}}{\Qlb{Fig6}}{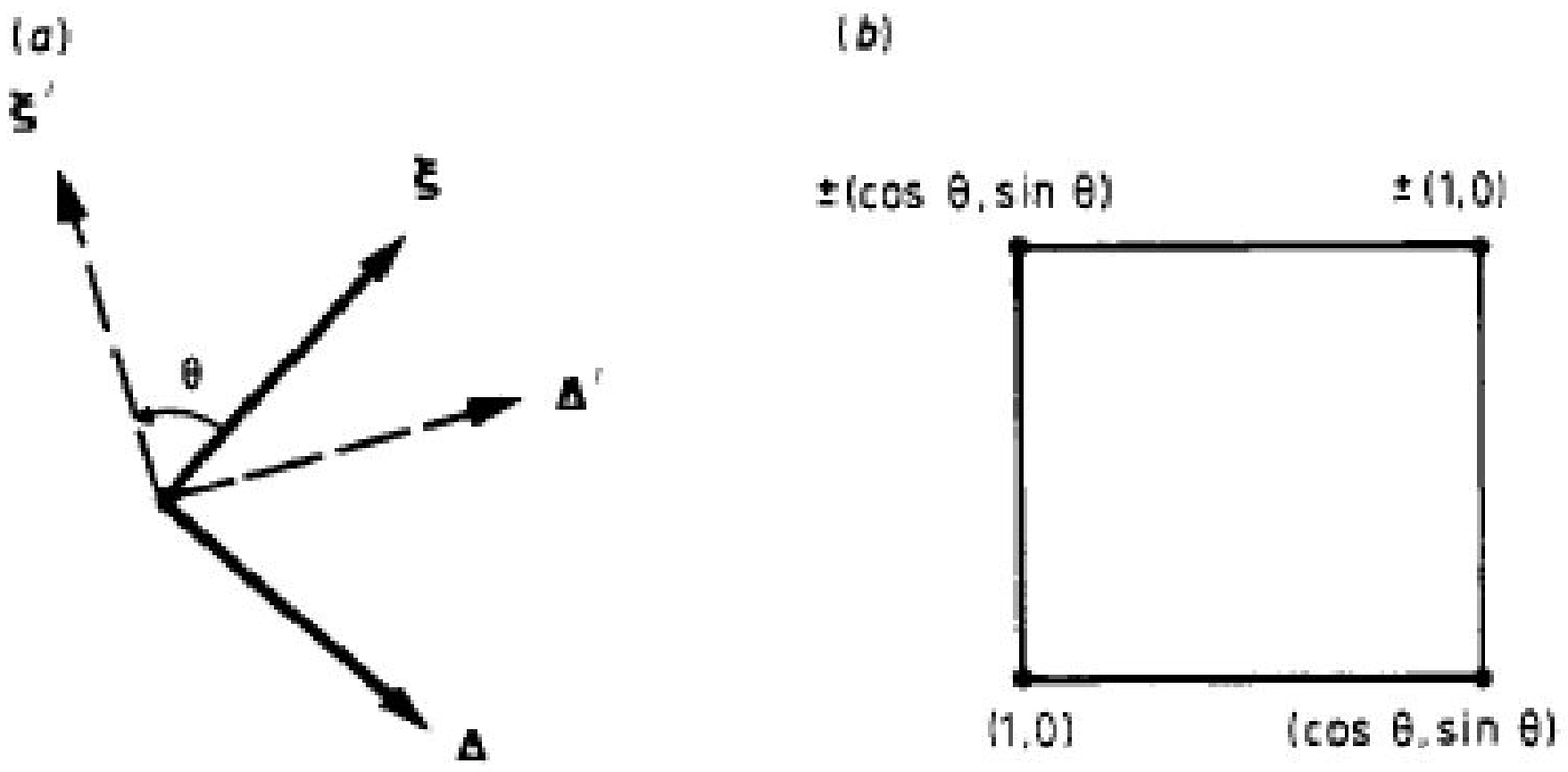}{%
\raisebox{-2.1205in}{\includegraphics[width=3.0in]{figure6.eps}}}

The degeneracy of the ground states of $H^{^{\prime }}$ may be explained
using the local $SU(2)$ symmetry of the Heisenberg Hamiltonian $H_{S}$. This
symmetry has been pointed out by Anderson \cite{Andseson 2}. Very recently
Affeck and co-workers \cite{Affleck} have studied the invariance under a
time-dependent $SU(2)$ gauge transformation, and discussed the equivalence
of some different mean-field theories. Here we wish to show that the
symmetry described in (\ref{Equation17}) is a sub-group of the local $SU(2)$%
, which preserves translational invariance and even parity that we imposed
in deriving the gap equations.

Consider a local $SU(2)$ gauge transformation, under which electron
operators at site $i$ transform as%
\begin{equation*}
c_{i\uparrow }^{+}\rightarrow \alpha _{i}c_{i\uparrow }^{+}+\beta
_{i}c_{i\downarrow }
\end{equation*}%
\begin{equation}
c_{i\downarrow }\rightarrow -\beta _{i}^{\ast }c_{i\uparrow }^{+}+\alpha
_{i}^{\ast }c_{i\downarrow }  \tag{19}  \label{Equation19}
\end{equation}%
where $\alpha _{i}\alpha _{i}^{\ast }+\beta _{i}\beta _{i}^{\ast }=1$. These
are the particle-hole \ transformations with spin conservation. $H_{S}$,
hence $H^{^{\prime }}$ at half-filling is invariant under these
transformations. Therefore all the states related by (\ref{Equation19}) are
degenerate. A set of the parameters $\widetilde{\Delta }_{\tau }$ and $\zeta
_{\tau }$, in our gap equations transforms to another set of the parameters
under (\ref{Equation19}). The transformations corresponding to the
degenerate state (\ref{Equation17}) derived from the gap equations are
represented in figure 6(b), for real values of $\widetilde{\Delta }_{\tau }$%
. There is also one to one correspondence to transformations of the d-wave
pairing state to a state with complex $\widetilde{\Delta }_{\tau }$ such as
the chiral state (\ref{Equaiton18c}) with $\alpha $ and $\beta $ complex.

Local $U(1)$ gauge symmetry is a sub-group of the local $SU(2)$ symmetry
since it is of the form%
\begin{equation*}
c_{i\sigma }\rightarrow \widetilde{c}_{_{i\sigma }}=e^{i\theta
_{i}}c_{i\sigma }
\end{equation*}%
A general choice of $\theta _{i}$ transforms the BCS pairing state (\ref%
{Equaiton3}) to a non-BCS-type state, which has the same energy as the
former. Since under such a transformation the Bloch coherence is lost, i.e.
a state of the form%
\begin{equation*}
\widetilde{c}_{k,\sigma }=\sum_{j}e^{ik\cdot R_{j}}\widetilde{c}_{j,\sigma
}=\sum_{j}e^{ik\cdot R_{j}+i\theta _{i}}c_{i\sigma }
\end{equation*}%
is no longer a coherent superposition of the original states. Yet we can
equally well pair $\widetilde{c}_{k,\uparrow }^{+}$ with $\widetilde{c}%
_{-k,\downarrow }^{+}$ , and the energy would be the same as if we pair $%
c_{k,\uparrow }^{+}$ with $c_{-k,\downarrow }^{+}$. Thus the apparent
coherent k-space pairing in the BCS wave-function (\ref{Equaiton3}) is
illusory. The absence of a coherent pairing order parameter at half-filling
as a consequence of the $U(1)$ gauge invariance has been stressed by
Baskaran and Anderson \cite{Baskaran2}.

It is important to realise that the states that are degenerate due to the $%
SU(2)$ gauge \ symmetry are the unprojected states $\left| \varphi
_{0}\right\rangle $ of (3), rather than the physical states $\left| \varphi
\right\rangle $ which obey the strict local constraint. How does a projected
state change under $SU(2)$? According to (\ref{Equation19}), a vacuum state
(empty state) at site $i$ must transform under $SU(2)$ as%
\begin{equation*}
\left| 0\right\rangle _{i}\rightarrow e^{i\theta _{i}}(\alpha _{i}^{\ast
}-\beta _{i}^{\ast }c_{i,\uparrow }^{+}c_{i,\downarrow }^{+})\left|
0\right\rangle _{i}
\end{equation*}%
This ensures that the vanishing of the state $c_{i,\sigma }\left|
0\right\rangle _{i}$ remains unchanged under the transformation as required
physically. However a singly occupied electron state transforms under $SU(2)$
as%
\begin{equation*}
c_{i,\sigma }^{+}\left| 0\right\rangle _{i}\rightarrow e^{i\theta
_{i}}c_{i,\sigma }^{+}\left| 0\right\rangle _{i}
\end{equation*}%
At half-filling, each site is singly occupied. Therefore any half-filled
state $\left| \varphi \right\rangle $ transforms into itself except for an
overall phase factor under the $SU(2)$ operator $\widehat{U}$:%
\begin{equation*}
\left| \varphi \right\rangle \rightarrow \widehat{U}\left| \varphi
\right\rangle =e^{i\theta }\left| \varphi \right\rangle ,\theta =\underset{i}%
{\sum }\theta _{i}
\end{equation*}

Although $\widehat{U}$ does not commute with the projection operator $P_{d}$%
, we observe for the half-filled state $\left| \varphi _{0}\right\rangle $%
\begin{equation*}
\widehat{U}P_{d}\left| \varphi \right\rangle _{0}=P_{d}\widehat{U}\left|
\varphi \right\rangle _{0}
\end{equation*}%
One way to see this is to notice that there is no empty site state also in $%
P_{d}\left| \varphi \right\rangle _{0}$. Thus we can rewrite%
\begin{equation*}
P_{d}\left| \varphi \right\rangle _{0}=\underset{i}{\prod }(n_{i,\uparrow
}-n_{i,\downarrow })^{2}\left| \varphi \right\rangle _{0}
\end{equation*}%
The $SU(2)$ transformations all commute with the operator $(n_{i,\uparrow
}-n_{i,\downarrow })^{2}$. Let $\left| \varphi \right\rangle =P_{d}\left|
\varphi \right\rangle _{0}$, and $\left| \varphi ^{^{\prime }}\right\rangle
_{0}=\widehat{U}\left| \varphi \right\rangle _{0}$, then%
\begin{equation*}
P_{d}\left| \varphi ^{^{\prime }}\right\rangle _{0}=P_{d}\widehat{U}\left|
\varphi \right\rangle _{0}=\widehat{U}\left| \varphi \right\rangle
=e^{i\theta }\left| \varphi \right\rangle =e^{i\theta }P_{d}\left| \varphi
\right\rangle _{0}
\end{equation*}%
This proves explicitly that the two states $\left| \varphi \right\rangle
_{0} $ and $\left| \varphi ^{^{\prime }}\right\rangle _{0}$ related by any
local $SU(2)$ gauge transformation correspond to the same state $\left|
\varphi \right\rangle $ except for a phase factor. Therefore all the states
in (\ref{Equation17}) correspond to the same physical state. The RVB ground
state is non-degradable.

There are also redundancies in the higher energy states in the fermion
representation. For instance, the state of Baskaran, Zou and Anderson \cite%
{Baskaran1} at half-filled is identical to the projected Fermi liquid state,
because the former transforms to the latter under (\ref{Equation19}) with $%
\alpha =1,\beta =0$ in one sublattice, and $\alpha =0,\beta =1$ in the
other. This equivalence was also pointed out by Yokoyama and Shiba \cite%
{Yokoyama1} in a different way.

We now comment on the local gauge symmetry in the original Hubbard model,
which in terms of the original fermion operator is%
\begin{equation*}
H_{H}=-t\underset{\left\langle i,j\right\rangle ,\sigma }{\sum }d_{i,\sigma
}^{+}d_{j,\sigma }+HC+U\underset{i}{\sum }d_{i,\uparrow }^{+}d_{i,\uparrow
}d_{i,\downarrow }^{+}d_{i,\downarrow }\bigskip
\end{equation*}
This Hamiltonian is not invariant under local gauge transformations with
respect to the operators $d_{i,\sigma }$. However, up to any finite order in 
$t/U$ there exists a canonical transformation, which eliminates the doubly
occupied sites \cite{Kohn}%
\begin{equation*}
H_{H}\rightarrow e^{iS}H_{H}e^{-iS}
\end{equation*}%
At half-filling, all the odd order terms in $t/U$ vanish. The Hamiltonian is
locally gauge invariant with respect to the electron operators in the new
representation, i.e. the $c_{i,\sigma }$ of (\ref{Equation1}) are Wannier
operators of the old representation, $d_{i,\sigma }$. Therefore the local
gauge symmetry holds to any finite order in perturbation theory in $t/U$.
This is the same as saying that the system has undergone a transition to a
Mott insulator \cite{Kohn}.

We have so far only examined the projected BCS-type trial wave-functions. It
is likely that the true ground state of the model Hamiltonian (\ref%
{Equation1}) at half-filled is the AF state \cite{Yokoyama2}. Recently,
Yokoyama and Shiba \cite{Yokoyama1},\cite{Yokoyama2} have studied a
projected Hartree-Fock-type AF state. Using VMC they found the energy per
site at half-filled to be $-0.642J$, slightly lower than $-0.636J$, the
value found by Gros \cite{Gros2} in the d-wave pairing state by using a
similar technique. However we may argue that the holes favour the pairing
state away from half-filled because of the gain in kinetic energy. We have
also applied the renormalised Hamiltonian approach to the AF states. Within
this approximation, we find that at the half-filled, the AF state has higher
energy than the RVB state (\ref{Equaiton2}), in contrast with the VMC
results. We present the derivation and the results in Appendix 1.

\section{Non-half-filled case}

\subsection{Ground state}

Firstly we examine the energy needed to create propagating Bloch states. The
simplest states for holes have the form%
\begin{equation*}
\left| \Phi _{i,\sigma }\right\rangle =c_{i,\sigma }\left| \varphi
\right\rangle
\end{equation*}%
which destroys a real electron at site $i$. We may also make a propagating
Bloch state for the hole of the form

\begin{equation}
\left| \Phi _{p,\sigma }\right\rangle =c_{p,\sigma }\left| \varphi
\right\rangle  \tag{20}  \label{Equation20}
\end{equation}%
A rigorous calculation is possible at half-filling.

We consider any translationally invariant spin singlet state $\left| \varphi
\right\rangle $ at half-filling. Let us denote by $\alpha $ the NN spin-spin
correlation in $\left| \varphi \right\rangle $, $\alpha =\left\langle
S_{i}\cdot S_{j}\right\rangle $. Then the magnetic energy loss of a hole in
the state $\Phi _{i,\sigma }$ is $-4\alpha J$, because the four bonds
connecting the site $i$ are mixing. Since the matrix of $H_{S}$ in (\ref%
{Equaiton2}) between any states $\left| \Phi _{i,\sigma }\right\rangle $ and 
$\left| \Phi _{j,\sigma }\right\rangle $ is diagonal, the moving hole state
of (\ref{Equation20}) has the same magnetic energy as in $\left| \Phi
_{i,\sigma }\right\rangle $.

The kinetic energy of the hole in (\ref{Equation20}) is given by%
\begin{equation*}
\left\langle H_{t}\right\rangle _{p}=2t\underset{\left\langle
i,j\right\rangle }{\sum }\left\langle n_{i,\sigma }n_{j,\sigma
}+S_{i}^{+}S_{j}^{-}\right\rangle _{\varphi }\exp (ip\cdot R_{ji})+HC
\end{equation*}%
where $\left\langle \text{ \ \ \ }\right\rangle _{\varphi }$ denotes the
expectation value in the half-filled state. Using the fact that $%
\left\langle n_{i}n_{j}\right\rangle _{\varphi }$ = 1, we get 
\begin{equation}
\left\langle H_{t}\right\rangle _{p}=t(1+4\alpha )(\cos (p_{x})+\cos (p_{y}))
\tag{21}  \label{Equation21}
\end{equation}%
Since $\alpha \simeq -0.33$ for the ground state, (\ref{Equation21}) gives a
band width for a Bloch hole of $0.64\left| t\right| $. The minimum energy to
remove an electron and create such a Bloch hole is $-0.32t-4\alpha J$.

The Bloch states are not the lowest energy states of the holes however. We
now apply the gap equations to study a system with a few pair of holes. The
energy to create a pair of holes is $-2\mu $ by the definition of the
chemical potential. Since the parameter $\widetilde{\mu }=0$ at the
half-filled, (\ref{Equation11}) gives the energy per hole to be $%
N_{S}^{-1}\left\langle \frac{\partial H^{^{\prime }}}{\partial \delta }%
\right\rangle _{0}$, a quantity related to the unprojected state $\left|
\varphi \right\rangle _{0}$ at half-filling. In the presence of holes, the
kinetic part of the Hamiltonian explicitly breaks the $SU(2)$ and $U(l)$
gauge symmetries, while the Heisenberg spin part remains invariant under
these symmetries.

Using (\ref{Equation4a},\ref{Equation4b}) and (\ref{Equation5}), for the
states $\left| \varphi \right\rangle _{0}$ described by (\ref{Equation17}),
the magnetic energy per hole is 
\begin{equation*}
\frac{1}{N_{S}}\frac{\partial g_{S}}{\partial \delta }\left\langle
H_{S}\right\rangle _{0}=6C^{2}J
\end{equation*}%
a value equivalent to the loss of four bonds in the spinspin correlations,
and it is the same for all these states as a consequence of the $SU(2)$
gauge invariance of the spin part of the Hamiltonian $H_{S}$. The kinetic
energy per hole in this case is given by%
\begin{equation*}
T=\frac{1}{N}\frac{\partial g_{t}}{\partial \delta }\left\langle
H_{t}\right\rangle _{0}=-4t(\zeta _{x}+\zeta _{y})
\end{equation*}%
$\zeta _{\tau }$, is the particle-hole correlation in $\left| \varphi
\right\rangle _{0}$ as defined in (\ref{Equation13}). When the holes are
introduced, a fraction ($g_{t}$) of this correlation becomes coherent in the
state $\left| \varphi \right\rangle $. Therefore the larger values of $\zeta
_{\tau }$, correspond to the lower kinetic energy of the holes. But $\zeta
_{\tau }$, are subject to (\ref{Equation17}). The kinetic energy can be
written in the following form by using (\ref{Equation17}) :%
\begin{equation*}
T=-4t(2C^{2}-\left| \widetilde{\Delta }_{x}+\widetilde{\Delta }_{y}\right|
^{2})^{1/2}sgn((\zeta _{x}+\zeta _{y}))
\end{equation*}%
Different parameters $\widetilde{\Delta }_{\tau }$, and $\zeta _{\tau }$,
describe different states with different energies upon doping. The above
hole kinetic energy expression immediately leads to the important conclusion
that the d-wave pairing state, where $\widetilde{\Delta }_{x}+\widetilde{%
\Delta }_{y}=0$ gives the best kinetic energy, which is 
\begin{equation*}
T=-4\sqrt{2}Ct=-2.71t
\end{equation*}%
Both the d-wave density matrix state (\ref{Equaiton18b}) and the chiral
state (\ref{Equaiton18c}) have zero kinetic energies, and are not favoured
upon doping. The kinetic energy for the d-wave pairing state in our
analytical approach is quite close to the VMC result, where it is found to
be $-2.55t$ for systems with 10\% holes \cite{Gros2}. It is also
substantially below the value found for a Bloch hole.

The introduction of some holes breaks the local gauge symmetry and causes
the ground state to be coherent. The stable lowest energy state upon doping
is the d-wave pairing state.

It is worthwhile remarking that in a Hubbard model it is the expectation
value of the kinetic energy which determines the integrated optical weight
associated with the charge carriers in the f-sum rule \cite{Baeriswyl} and
in the present case this optical weight is proportional to the number of
holes with an optical mass determined from the proportionality constant of
order $at^{-1}$ (where $a$ is the lattice constant).

The gap equations for the finite hole concentrations can be solved
numerically. Here we shall consider only the most stable state-the d-wave
pairing state. In this case, we set $\widetilde{\Delta }_{x}=-\widetilde{%
\Delta }_{y}=\widetilde{\Delta }$, and $\zeta _{x}=\zeta _{y}=\zeta $. The
four equations (\ref{Equation13}) reduce to two because of the symmetry
between the $x$ and $y$ components. These two equations uniquely determine $%
\widetilde{\Delta }$ and $\zeta $ for the fixed values of $t$, $J$ and $%
\widetilde{\mu }$. The numerical results of the gap equations are plotted in
figure 7 for $\widetilde{\Delta }$ as a function of the hole concentration.%
\FRAME{ftbpFU}{2.9499in}{2.6221in}{0pt}{\Qcb{\textit{Variational parameter }$%
\widetilde{\Delta }$\textit{\ and superconducting order parameter }$\Delta
_{SC}$\textit{\ as functions of the hole concentration }$\protect\delta $%
\textit{\ for a choice of }$t/J=5$\textit{\ in the d-wave pairing state. }}}{%
\Qlb{Fig7}}{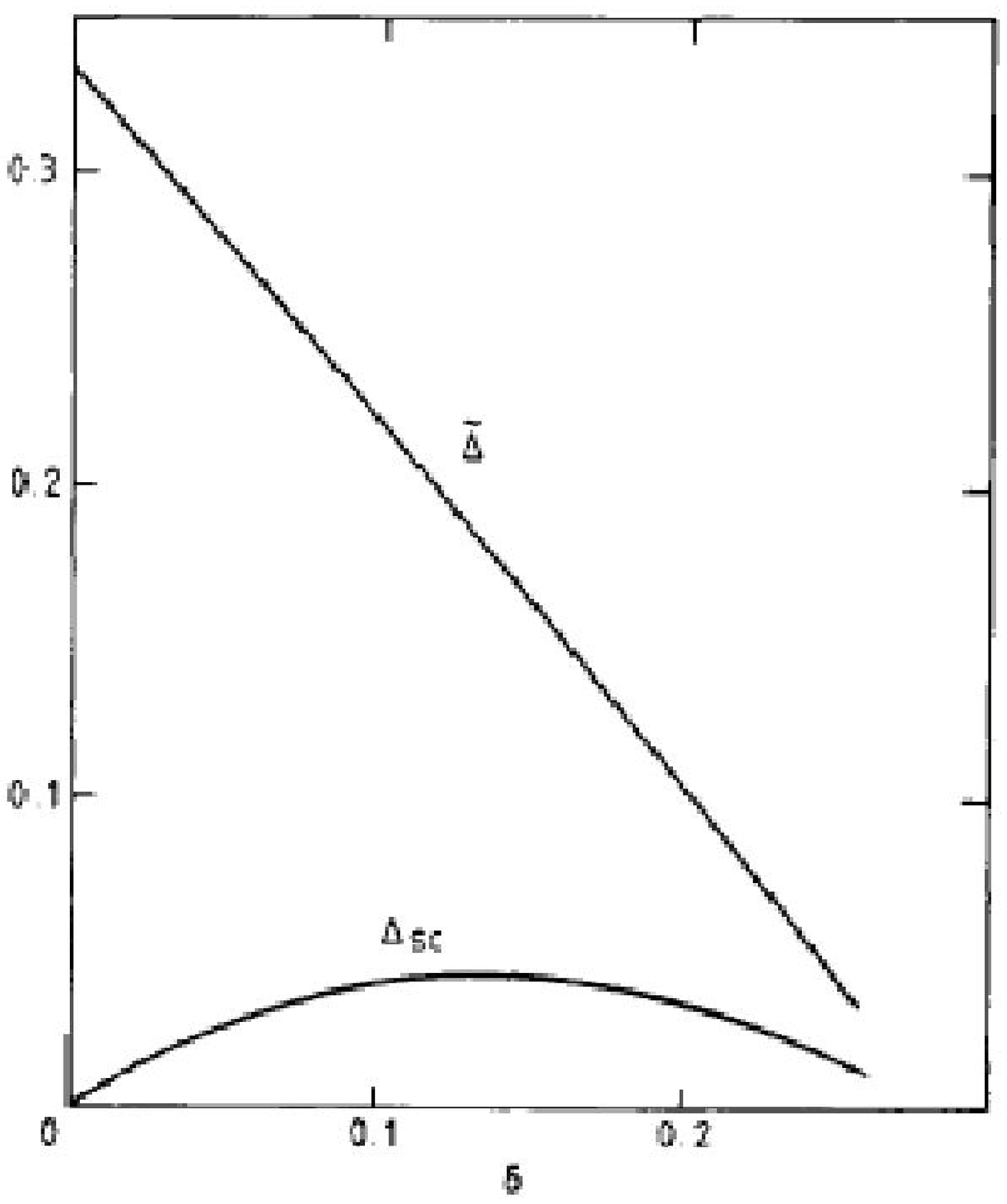}{\raisebox{-2.6221in}{\includegraphics[width=3.0in]{fig7-2.eps}}}

We now discuss the superconducting order parameter. As mentioned in
section3, $\widetilde{\Delta }$ is not the order parameter. The
superconducting order parameter is 
\begin{equation*}
\Delta _{SC}(R_{ij})=\left\langle c_{i,\uparrow }^{+}c_{j,\downarrow
}^{+}-c_{i,\downarrow }^{+}c_{j,\uparrow }^{+}\right\rangle
\end{equation*}%
an expectation value in the projected state (\ref{Equaiton2}). This quantity
describes the Cooper pairing in a real space representation. We shall adopt
the Gutzwiller method to calculate this quantity. In analogy to the
derivation for the hopping energy in section2 we find that the
nearest-neighbour sites $i$ and $j$%
\begin{equation*}
\left\langle c_{i,\uparrow }^{+}c_{j,\downarrow }^{+}\right\rangle
=g_{t}\left\langle c_{i,\uparrow }^{+}c_{j,\downarrow }^{+}\right\rangle _{0}
\end{equation*}%
Therefore for nearest-neighbour sites, the order parameter is related to the
variational parameter a in the gap equations by%
\begin{equation}
\Delta _{SC}=g_{t}\widetilde{\Delta }  \tag{22}  \label{Equation22}
\end{equation}%
The value of $\Delta _{SC}$ as a function of $\delta $ is plotted in figure
7 in comparison with $\widetilde{\Delta }$. $\Delta _{SC}$ vanishes linearly
near $\delta =0$. $\Delta _{SC}$ found in our theory is in good agreement
with the Monte Carlo results [\cite{Gros2}]. The absence of the
superconducting order parameter at half-filled obtained from (\ref%
{Equation22}) agrees with the discussion in section4 from the viewpoint of
the local gauge symmetry.

The kinetic energy of holes in the AF state is found to be quite high in our
analytic approach (see Appendix 1). However, the Gutzwiller approximation we
adopted is too rough to determine whether a RVB or AF state has lower
energy. Numerical results of VMC \cite{Gros2}, \cite{Yokoyama1}, \cite%
{Yokoyama2} suggest both the spin-spin correlation energy and the kinetic
energy of the holes between the d-wave pairing state and the AF states are
very close. The question which state is more favourable in energy remains
unresolved.

\subsection{Excited states and finite temperatures}

We begin by examining the spin degrees of freedom in half-filled and near
half-filled cases. An excited state can be created by applying the spin
raising operator to a specific site to obtain%
\begin{equation*}
\left| \Psi _{i,+}\right\rangle =S_{i}^{+}P_{d}\left| \varphi \right\rangle
_{0}
\end{equation*}%
We can commute $S_{i}^{+}$ with $P_{d}$, to obtain%
\begin{eqnarray*}
\left| \Psi _{i,+}\right\rangle &=&P_{d}S_{i}^{+}\left| \varphi
\right\rangle _{0} \\
&=&\underset{p,p^{^{\prime }}}{\sum }e^{i(p-p^{^{\prime }})\cdot
R_{i}}P_{d}c_{p,\uparrow }^{+}c_{p^{^{\prime }},\downarrow }\left| \varphi
\right\rangle _{0}
\end{eqnarray*}%
This state is therefore a superposition of two independent quasi-particle
states similar to a metal where the low energy excitations are made up of
superpositions of electron and hole states. The quasi-particle states
(spinons) can be defined by%
\begin{equation}
\left| \Psi _{p,\uparrow }\right\rangle =P_{d}c_{p,\uparrow }^{+}\underset{%
k\neq p}{\prod }(u_{k}+v_{k}c_{k,\uparrow }^{+}c_{-k,\downarrow }^{+})\left|
0\right\rangle  \tag{23}  \label{Equation23}
\end{equation}%
The quasi-particle energy $\overline{E_{p}}$, is defined to be the
difference of the expectation values of $K=H-\mu N$ in the state $\left|
\Psi _{p,\uparrow }\right\rangle $ and in the ground state $\left| \varphi
\right\rangle $. We use the Gutzwiller method to calculate the energy of the
state (\ref{Equation23}). The energy difference between the two states
contains two parts. One is due to the changes of the renormalization factors 
$g_{t}$ and $g_{s}$ , the other comes from the change of the wave-function
itself. The former just cancels exactly the second term in $\mu $ in (\ref%
{Equation11}). Using (\ref{Equation7}) to calculate the energy difference
due to the wave-function change, we get%
\begin{eqnarray*}
\overline{E_{p}} &=&(1-2v_{p}^{2})(g_{t}\varepsilon
_{p}+N_{S}^{-1}\sum_{k}V_{k-p}v_{k}^{2}-\widetilde{\mu }) \\
&&+2u_{p}v_{p}N_{S}^{-1}\sum_{k}V_{k-p}u_{k}^{\ast }v_{k}
\end{eqnarray*}%
Applying the gap equations to simplify the expression, we obtain%
\begin{equation}
\overline{E_{p}}=\frac{3}{4}g_{S}JE_{p}  \tag{24}  \label{Equation24}
\end{equation}%
Note this energy is independent of the local $SU(2)$ gauge and does not
depend on the particular fermion representation. At the pseudo-Fermi
surface, where by definition $\zeta _{p}=0$, we have $E_{p}=\left| 
\widetilde{\Delta }_{p}\right| $. Since the state (\ref{Equation23}) breaks
a pair of electrons, $\overline{E_{p}}$ describes the binding energy of the
pair at the pseudo-Fermi surface. The excitation energy depends on the
particular RVB. For the Fermi liquid state ($u_{k}v_{k}=0$) then $\overline{%
E_{p}}=0$ over the whole pseudo-Fermi surface. However in the ground RVB
state it vanishes only at four points, e.g. when $n=1$, $\overline{E_{p}}%
=E_{p}=0$, if $(p_{x},p_{y})=(\pm \pi /2,\pm \pi /2)$ and the density of
spinon states at low energies is%
\begin{equation*}
N(\overline{E})=\sum_{p}\delta (\overline{E}-\overline{E_{p}})\rightarrow 
\frac{2\overline{E}}{\pi (\frac{3}{4}g_{S}J)^{2}},\text{ \ as \ }\overline{E}%
\rightarrow 0
\end{equation*}

We turn now to a brief discussion of the system at finite temperature. The
extension of the mean-field gap equations to finite $T$ is not so
straightforward. The existence of a finite $\widetilde{\Delta }_{k}$ is
controlled by the energy scale of $\overline{E_{k}}$\ i.e. by $J$ in (\ref%
{Equation24}). On the other hand if we consider the limit $\delta \ll 1$
there is a very small energy scale $\propto \delta t$ which controls the
definition of a coherent gauge. In other words it is only the kinetic energy
which allows us to determine the gauge uniquely and at temperatures $J\gg
T\gg \delta t$, the gauge coherence will be lost. Yet in this temperature
range the magnetic coherence survives since as we have stressed earlier this
is independent of the choice of gauge on each site. The properties of the
system in this temperature region are clearly very different from Fermi
liquid behaviour as Anderson has stressed and these two energy scales should
correspond to his `holon' and `spinon' energy scales respectively. The
thermopower should obey the Heikes formula \cite{Mott} and we can expect
only a low mobility of the holes. However, a more detailed study of this
regime is required.

\section{Discussion}

We have used a variational method to study a projected BCS trial
wave-function for the square lattice effective Hamiltonian. Using a
Gutzwiller approximation to treat the effect of the projection operator, we
obtained a renormalised Hamiltonian in which the projection operator is
replaced by renormalisation factors. This approximation is shown to be in
good agreement with numerical Monte Carlo calculations for such projected
wave-functions. In this mean-field approximation both particle-particle and
particle-hole pairing amplitudes must be included. The fermion
representation for the ground state at the half-filled band is highly
redundant, due to a local $SU(2)$ invariance at exactly half-filling. This
redundancy is reflected in an apparent degeneracy of the BCS trial
wave-function before projection. Doping destroys the local $SU(2)$
invariance and splits these degenerate states, and we find that the stable
state upon doping is the d-wave pairing RVB state. In this RVB state, \
electrons are paired even at half-filling and it costs an energy of order $J$
to break a pair. These pre-existing electron pairs lead to a non-zero
superconductivity amplitude upon doping, and the magnitude of this
superconducting amplitude or order parameter is shown to be proportional to
the hole concentration $\delta $ when $\delta $ is small. The elementary
excitations at half-filling are the projected BCS quasi-particle states or
spinons, with four point zeros on the pseudo-Fermi surface.

Our analytic approach can also be applied to 1D and systems with
dimensionality $d\geq 3$. We find that lowest energy state in 1D is the
projected Fermi liquid RVB without electron pairing, as shown in Appendix 2.
Our theory predicts no superconductivity in a 1D RVB. For large $d$, the
energy per bond in the RVB pairing state is proportional to $1/d$, reduces
relative to an AF. So the pairing state is particularly favourable in 2D.
The precise form of the 2D phase diagrams which depends sensitively on the
relative energies of the AF and d-wave RVB states as a function of $\delta $
is too subtle a question to be settled by the approximation we use here.

There are many questions that require further investigation such as the
exact relationship between the discussion here in terms of phase coherence
among the paired electrons and Anderson's 'holon' \cite{Kivelson} concept or
the nature of the high-temperature phase where this phase coherence is lost
but strong magnetic correlations remain and presumably do not lead to a
Fermi liquid that is the usual description of a normal state.

\section{\protect\bigskip Acknowledgements}

We would like to thank D Poilblanc, R Joynt, M Roos, T K Lee, G Kotliar for
many useful discussions. Financial support from the Swiss Nationalfonds is
gratefully acknowledged.

\section{Appendix}

\subsection{The projected spin-density-wave state}

In this Appendix, we use the renormalised Hamiltonian approach to study the
projected spin-density-wave state for effective Hamiltonian (\ref{Equation1}%
). That state was proposed and studied using VMC by Yokoyama and Shiba \cite%
{Yokoyama1}. The generalisation of the Gutzwiller method to the
antiferromagnetic states for the hopping process was formulated by Ogawa and
co-workers \cite{Ogawa} .

The projected spin-density-wave state \cite{Yokoyama1}, \cite{Yokoyama2} is%
\begin{equation}
\left| \psi \right\rangle =P_{d}\left| \psi _{0}\right\rangle  \tag{A1.1}
\label{A1.1}
\end{equation}%
\begin{equation}
\left| \psi _{0}\right\rangle =\underset{k\sigma }{\prod }(u_{k}c_{k,\sigma
}^{+}+\sigma v_{k}c_{k+Q,\sigma }^{+})\left| 0\right\rangle  \tag{A1.2}
\label{A1.2}
\end{equation}%
where $k$ runs over the Fermi sea, $Q=\pi /a(1,1)$, and%
\begin{eqnarray*}
u_{k} &=&[(1+\cos \theta _{k})/2]^{1/2} \\
v_{k} &=&[(1-\cos \theta _{k})/2]^{1/2} \\
\cos \theta _{k} &=&\gamma _{k}/(\Delta _{AF}^{2}+\gamma _{k}^{2})^{1/2}
\end{eqnarray*}%
$\Delta _{AF}$ is a variational parameter, and $\gamma _{k}$ is given by (%
\ref{Equaiton18c}).

In a study of the expectation value in the state (\ref{A1.1}), we use the
Gutzwiller approximation to replace the projection operator by
renormalisation factors. In analogy to the analysis we discussed in
section2, we find that%
\begin{eqnarray*}
g_{t} &=&\frac{1-n}{1-2n_{\uparrow }n_{\downarrow }/n} \\
g_{S} &=&(1-2n_{\uparrow }n_{\downarrow }/n)^{2}
\end{eqnarray*}%
where $n_{\uparrow }$ and $n_{\downarrow }$ are the spin-up and spin-down
electron occupation number of state $\left| \psi _{0}\right\rangle $ in one
sublattice respectively. The renormalisation factors reduce to (\ref%
{Equation4a},\ref{Equation4b}) in the case $n_{\uparrow }=n_{\downarrow }$ ,
and the form for $g_{t}$ agrees with \cite{Ogawa}.

Within this scheme, we obtain the energy per site at half-filling%
\begin{equation*}
w=-2J(a^{2}+6b^{2})/(1+a^{2})^{2}
\end{equation*}%
where%
\begin{eqnarray*}
a &=&N_{S}^{-1}\underset{k}{\sum }\Delta _{AF}/(\Delta _{AF}^{2}+\gamma
_{k}^{2})^{1/2} \\
b &=&(8N_{S})^{-1}\underset{k}{\sum }\gamma _{k}^{2}/(\Delta
_{AF}^{2}+\gamma _{k}^{2})^{1/2}
\end{eqnarray*}%
The spin-spin correlation $\left\langle S_{i}\cdot S_{j}\right\rangle =\frac{%
1}{2}w$, and the staggered magnetisation is%
\begin{equation*}
M_{S}=\overline{n_{\uparrow }}-\overline{n_{\downarrow }}=2a/(1+a^{2})
\end{equation*}%
with $\overline{n_{\sigma }}$, the occupation number in the state $\left|
\psi \right\rangle $. $\left\langle S_{i}\cdot S_{j}\right\rangle $ and $%
M_{S}$, are plotted in figure A1 as functions of $\Delta _{AF}$, Figure A1
also shows the VMC calculations \cite{Yokoyama1}. The case $\Delta _{AF}=0$
corresponds to the projected Fermi liquid state, while $\Delta
_{AF}\rightarrow \infty $ corresponds to the N$\overset{^{\prime }}{e}$el
state. The results agree well for large values of $\Delta _{AF}$, but there
are substantial deviations for small $\Delta _{AF}$.\FRAME{ftbpFU}{3.0805in}{%
2.2969in}{0pt}{\Qcb{\textit{Spin-spin correlation and staggered
magnetisation }$M_{S}$\textit{\ as functions of }$\Delta _{AF}$\textit{\ in
the projected spin-density-wave state. The full curves are the results of
the renormalised Hamiltonian approach, and the broken curves are the VMC
results (extrapolated \ to the infinite systems) by Yokoyama and Shiba %
\protect\cite{Yokoyama1}, \protect\cite{Yokoyama2}. }}}{\Qlb{FigA.1}}{%
figurea1.eps}{\raisebox{-2.2969in}{\includegraphics[width=3.0in]{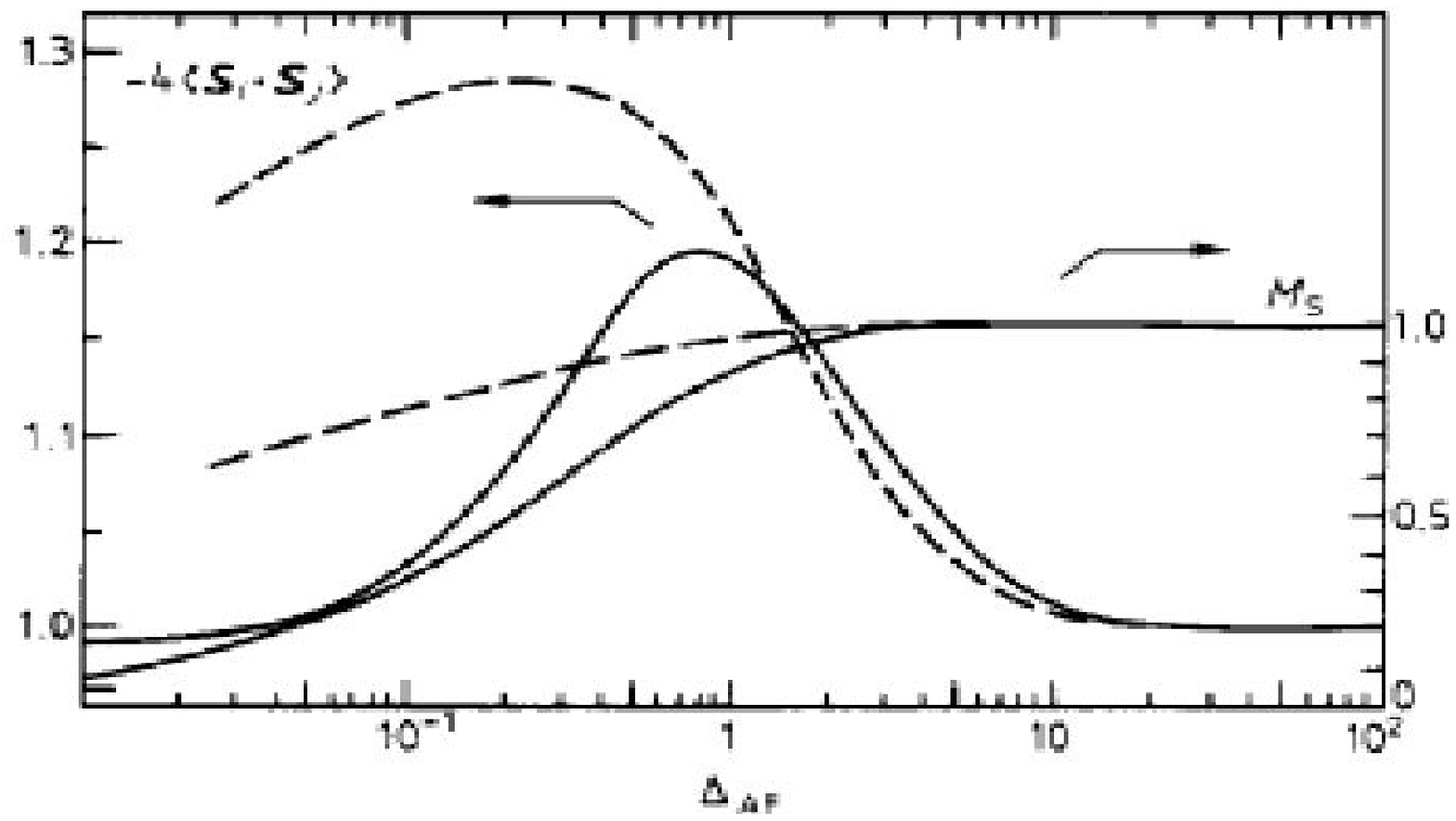}}}

The kinetic energy per hole in our theory is%
\begin{equation*}
T=-16tb/(1+a^{2})
\end{equation*}%
For the optimal value of $\Delta _{AF}(\sim 0.9)$, $T=-2.16t$, substantially
higher than that in the d-wave pairing state. Note that this value is also
higher than that found in the VMC calculation \cite{Yokoyama1}, where the
optimal AAF is found to be much smaller.

\subsection{RVB in a 1D system}

The renormalised Hamiltonian approach can be straightforwardly applied to
the model (\ref{Equation1}) in 1D. Using the projected BCS wave-function (%
\ref{Equaiton2}), and the same technique for 2D, we have found that the RVB
ground state \ at the half-filling is described by an equation between $%
\zeta _{x}$, and $\widetilde{\Delta }_{x}$ (defined in (\ref{Equation12})-(%
\ref{Equation13})):%
\begin{equation}
\zeta _{x}^{2}+\left| \widetilde{\Delta }_{x}\right| ^{2}=C_{1}^{2} 
\tag{A2.1}  \label{A2.1}
\end{equation}%
with%
\begin{equation*}
C_{1}=(2N_{S})^{-1}\underset{k}{\sum }\left| \cos (k_{x})\right| =2/\pi ^{2}
\end{equation*}%
(\ref{A2.1}) is parallel to (\ref{Equation17}) in 2D. Similar to the 2D
case, different parameters in (\ref{A2.1}) are related to each other under
the $SU(2)$ gauge transformation, and correspond to the same physical state.
This state is described by the projected Fermi liquid state, where $%
\widetilde{\Delta }_{x}=0$, $\zeta _{x}=C_{1}$. Earlier Monte Carlo
calculations \cite{Gros3} and more recent exact calculation \cite{Gebhard}
with this wave-function have shown that the energy of this state is
extremely close to the exact solution \cite{Bethet}. The energy per site in
our analytic mean-field approach is $-6/\pi ^{2}$. This value deviates by
about 37\% from the true result \cite{Gros3}, \cite{Gebhard}. This
quantitative discrepancy however is not surprising, because the Gutzwiller
approximation is poor in 1D.

In parallel to the discussion in section4, we can study the system with some
holes. We found that the stable lowest-energy state corresponds to $%
\widetilde{\Delta }_{x}=0$. Introducing the finite value of $\widetilde{%
\Delta }_{x}$, the system loses kinetic energy. Therefore we expect there is
no electron pairing and nosuperconductivity in this 1D RVB.

\end{document}